\documentclass[twocolumn]{jpsj3}
\usepackage{txfonts}

\title{
Magnetic Order and Charge Disproportionation in a Spin-ice type Kondo Lattice Model: Large Scale Monte Carlo Study
}

\author{
Hiroaki Ishizuka$^1$\thanks{E-mail: ishizuka@aion.t.u-tokyo.ac.jp},
Masafumi Udagawa$^{1,2}$,
and Yukitoshi Motome$^1$
}

\inst{
$^1$Department of Applied Physics, University of Tokyo, Tokyo 113-8656, Japan \\
$^2$Max Plank Institute for the Physics of Complex Systems, D-01187 Dresden, Germany
} 

\abst{
Phase diagram of a spin-ice type Kondo lattice model, potentially relevant to metallic
pyrochlore oxides, is obtained by the Monte Carlo simulation implementing the polynomial
expansion technique up to the system size with 2048 sites.
We identified a new 32-sublattice magnetic phase with concomitant charge 
disproportionation, along with other phases such as two-in two-out
and all-in/all-out orders.
The spin and charge pattern can be switched by external magnetic field to
a different one accompanied by a half magnetization plateau.
}

\kword{
spin ice, Kondo lattice model, pyrochlore oxides, polynomial expansion Monte Carlo method, charge disproportionation, magnetization plateau
}

\begin{document}
\maketitle

Geometrical frustration offers a fertile ground for studying fascinating phenomena in strongly correlated systems~\cite{Diep,Lacroix}. 
Competing interactions under the frustration often lead to extensive number of energetically-degenerate states.
Even a small perturbation to the degeneracy can result in dramatic effects, such as phase transitions and colossal responses to external fields.  
This sensitivity has stimulated intensive studies of competing orders and fluctuations in the geometrically-frustrated systems.

\begin{figure}
   \begin{center}
   \includegraphics[width=3.20in]{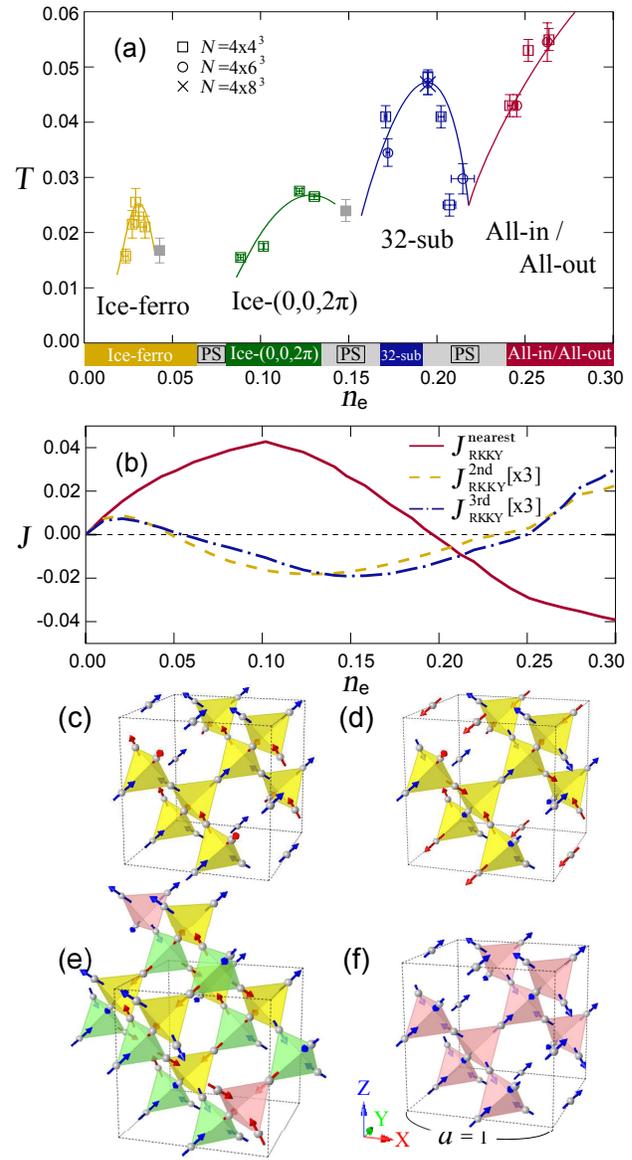}
   \end{center}
   \caption{
   (color online).
   (a) Phase diagram of the model (\ref{eq:H}) at $J=2$.
   The open symbols show $T_c$ for the four phases, while the closed ones are for other complex orders~\cite{note_gray}. 
   The lines are guides for the eyes.
   The bottom stripe shows the ground-state phase diagram obtained by comparing the energy of four phases~\cite{note_gsdiagram}.  
   PS indicates phase separation. 
   (b) RKKY interactions for the nearest-, second-, and third-neighbor spins calculated by using the second-order perturbation in terms of $J$ for the model (\ref{eq:H}). 
   The latter two are multiplied by factor $3$ for clarity.
   (c)-(f) show spin configurations for (c) ice-ferro, (d) ice-$(0,0,2\pi)$, (e) 32-sublattice, and (f) all-in/all-out order, respectively.
   }
   \label{fig:diagram}
\end{figure}

A representative example of frustrated systems is the spin ice~\cite{Harris1997,Ramirez1999,GingrasPreprint}.
In these systems, spins with strong Ising-type anisotropy along the
sublattice-dependent local $\langle 111\rangle$ direction reside on the pyrochlore lattice, which consists of corner-sharing tetrahedra,
as shown in Figs.~\ref{fig:diagram}(c)-(f).  
The nearest-neighbor (NN) interaction between the Ising spins is dominantly ferromagnetic (FM), favoring the two-in two-out
configuration in each tetrahedron.
This two-in two-out constraint, called the ice rule, leads to macroscopic degeneracy  and prevents the system from
ordering~\cite{Pauling1935}.
The situation offers an interesting playground for controlling the peculiar magnetic states by external magnetic
field~\cite{GingrasPreprint,Harris1998}.
In addition, it was theoretically predicted that the magnetic states are drastically changed by the sign and strength of exchange
interactions~\cite{Melko2004}. 
Namely, when the dominant interaction changes from FM to antiferromagnetic (AFM), all-in or all-out spin configuration becomes
favored in each tetrahedron, resulting in a long-range order of alternating
all-in and all-out tetrahedra.
Although such phase transition is intriguing, the sign of interaction is usually inherent
to each material and not flexibly tunable. 

In the present study, we explore another route to tuning the ice-type frustrated systems. 
We here consider a coupling of the spin-ice type moments to itinerant electrons.
The kinetic motion of electrons causes an effective interaction between the localized
moments, called the Ruderman-Kittel-Kasuya-Yosida (RKKY) interaction~\cite{Ruderman1954}. 
The effective interaction is, in general, long-ranged and oscillating (can be both FM and
AFM), sensitively depends on the electron density and band structure. 
Therefore, the system has great potential to tune the phase competition by controlling
carrier doping as well as external pressure which modifies the band structure.

Potential realization of such situation is brought by metallic pyrochlore oxides, such as
{\it R}$_2$Mo$_2$O$_7$ and {\it R}$_2$Ir$_2$O$_7$ ({\it R} is rare-earth), in which
Ising-like rare-earth $f$ moments  couple with itinerant $d$ electrons. 
The interplay between spin and charge is believed to play an essential role in various
peculiar features in these systems, such as the anomalous Hall
effect~\cite{Taguchi2001,Machida2010} and resistivity minimum~\cite{Nakatsuji2006}.

In this Letter, we elucidate thermodynamic properties of an Ising-spin Kondo lattice model on the pyrochlore lattice by employing an unbiased numerical simulation. 
A related study was performed for a spin-only model with the long-range RKKY interaction~\cite{Ikeda2008}.
However, the RKKY form is based on a second-order perturbation in terms of the spin-charge
coupling with assuming a simple isotropic Fermi surface, neglecting the actual band
structure.
In the present study, we solve the problem in the nonperturbative region by a large scale
Monte Carlo (MC) simulation which includes the electronic degree of freedom explicitly,
and reveal a new spin-charge coupled phase and its switching by external magnetic field.

We consider a Kondo lattice model with Ising spins on a pyrochlore lattice, whose
Hamiltonian is given by
\begin{eqnarray}
H = -t \! \sum_{\langle i,j \rangle, \sigma} \! ( c^\dagger_{i\sigma} c_{j\sigma} +
\text{H.c.} ) -J \sum_{i} {\bf S}_i \cdot {\boldsymbol \sigma_i}. \label{eq:H}
\end{eqnarray}
The first term is hopping of itinerant electrons, where $c_{i\sigma}$
($c^\dagger_{i\sigma}$) is the annihilation (creation) operator of an itinerant electron
with spin $\sigma= \uparrow, \downarrow$ at $i$th site, and $t$ is the transfer integral.
The sum $\langle i,j \rangle$ is taken over NN sites on the pyrochlore lattice.
The second term is the onsite interaction between localized spins and itinerant electrons, where
${\bf S}_i$ and ${\boldsymbol \sigma}_i$ represent the localized Ising spin and itinerant
electron spin at $i$th site, respectively ($|{\bf S}_i|=1$), and $J$ is the
coupling constant (the sign of $J$ does not matter in the present model as the localized spins are classical). 
The anisotropy axis of Ising spin is given along the local $\langle 111\rangle$ direction at each site,
i.e., along the line connecting the centers of two tetrahedra which the spin belongs to.
Hereafter, we take $t=1$ as the unit of energy, the lattice constant of cubic unit cell
$a = 1$ [see Fig.~\ref{fig:diagram}(f)], and the Boltzmann constant $k_{\rm B} = 1$.

The model in eq.~(\ref{eq:H}) was studied by the authors by using a cluster dynamical
mean-field theory with a focus on transport properties~\cite{Udagawa2012}.
In the previous study, they calculated the phase diagram with limiting 
the magnetic unit cell to a single tetrahedron.
It is highly desired to make a complete phase diagram by considering 
longer-range spatial correlations.

Here we study thermodynamic properties of the model by an unbiased MC
simulation. The partition function of the present model is given by 
$Z = {\rm Tr}_{\{ {\bf S}_i \}} {\rm Tr}_{\{ c_{i\sigma}, c_{i\sigma}^\dagger \}} \exp[- \{H(\{ {\bf S}_i \})-\mu \hat{N_c}\}/T]$ ($\hat{N_c}$
is the fermion number operator and $\mu$ is the chemical potential). 
The former trace is taken over the Ising spin degree of freedom, which is handled by the MC sampling using the standard Metropolis method with single spin flip dynamics.
To obtain the MC weight, the latter trace for itinerant electrons is calculated by the following two methods. 
One is the exact diagonalization (ED)~\cite{Yunoki1998}, which is applicable only to small system sizes because the computational amount increases rapidly as a function of the system size $N$ as $O(N^{3})$. 
The other method is the polynomial expansion method (PEM)~\cite{Motome1999}, which reduces the computational amount to $O(N^{2} \log N)$~\cite{note_on_PEM}, enabling us to access larger system sizes. 
Prior to calculations, we tested the convergence of PEM in terms of the order of Chebyshev polynomials $m$~\cite{Ishizuka_preprint}.
In relatively higher density region $n_e \gtrsim 0.15$, we concluded that $m=40$ gives good convergence ($n_e = \sum_{i\sigma} \langle c_{i\sigma}^\dagger c_{i\sigma} \rangle / N$).
On the other hand, for  $n_e \lesssim 0.15$, the convergence appears to be much worse, and hence we employed ED.
Most of the calculations were conducted up to $N=4\times 6^3$ by PEM and
$N=4\times 4^3$ by ED for typically more than 3000 MC steps.
In addition, calculations for $N=4\times 8^3$ were also done 
at a filling in the 32-sublattice ordered region.
Note that this is a state-of-the-art calculation; one MC step for $N=4\times 8^3$ takes about 50 seconds by using 1024 CPU cores in the System B (SGI Altix ICE 8400EX) at ISSP supercomputer center.

Figure~\ref{fig:diagram}(a) shows the phase diagram of the model (\ref{eq:H}) at $J=2$
obtained by the MC simulation.
There are four dominant phases at low $T$ in the calculated density region
$0 \leq n_e \leq 0.3$. 
The open symbols show the critical temperatures $T_c$ for each phase determined from the
inflection point or sharp jump of $T$ dependence of the order parameter at each system size
(see Fig.~\ref{fig:structure} below).
The order parameter is defined by $M_{\bf q} = [S^{(\alpha)}_{{\bf q} \ {\rm max}}/N_{\rm t}]^{1/2}$, where
$S^{(\alpha)}_{{\bf q} \ {\rm max}}$ is the maximum component of sublattice-dependent
spin structure factor, $ S^{(\alpha)}_{\bf q} = \sum_{i, j \in \alpha} \langle {\bf S}_i \cdot {\bf S}_j \rangle
\exp\left[ {\rm i} {\bf q}\cdot\left({{\bf r}_i - {\bf r}_j}\right)\right] /N_{\rm t}$
($N_{\rm t} = N/4$ is the number of tetrahedra, $\alpha$ denotes four sublattices in a primitive
cell, ${\bf r}_i$ is the position of $i$th site, and $\bf q$ is the wave number).
The inflection point gives a good estimate of $T_c$, since, in general, the order parameter 
monotonically goes to zero (a nonzero value) as increasing $N$ in the higher (lower) $T$ region;
the system size dependence of $T_c$ is reasonably small, while we do not have
enough sets of data to perform the systematic finite-size scaling analysis~\cite{note_Tc}.
The ordering pattern below $T_c$ is determined from the characteristic wave number ${\bf q}$
together with the local spin correlations within tetrahedra introduced below.

In the low density region $n_e \lesssim 0.04$, the system develops the ${\bf q}=0$ two-in two-out order below $T_c \lesssim 0.025$ [Fig.~\ref{fig:diagram}(c)].
We call this phase the ice-ferro phase hereafter.
While increasing the electron density, a different ordering develops in the region of $0.08 \lesssim n_e \lesssim 0.15$~\cite{note_gray}.
The ordering structure is of layer type with ${\bf q}=(0,0,2\pi)$; every tetrahedron retains two-in two-out ice-rule configuration, while the net moments of tetrahedra form a collinear layer-type ordering [Fig.~\ref{fig:diagram}(d)].
We call this the ice-$(0,0,2\pi)$ phase.
The ordering pattern is the same as that found in a spin ice model including the long-range dipole-dipole interaction~\cite{Melko2004}.
On the other hand, in the high density region $n_e \gtrsim 0.22$, the system exhibits the ${\bf q}=0$ all-in/all-out order [Fig.~\ref{fig:diagram}(f)].

In the intermediate region of $0.15 \lesssim n_e \lesssim 0.22$, another complex magnetic structure appears.
The magnetic ordering has a 32-sublattice structure characterized by the wave vector ${\bf q}=(\pi,\pi,\pi)$~\cite{note_wavenumber}. 
The complicated structure can be viewed as a collection of one-dimensional chains with ``in-in-out-out" spin configuration [see Fig.~\ref{fig:diagram}(e)];
namely, all the second-neighbor spins along the chains are AFM.
Surprisingly, such chain AFM is free from frustration in the pyrochlore lattice, uniquely selecting the 32-sublattice ordering.
In terms of tetrahedra, the 32-sublattice order consists of a periodic arrangement of 
all $2^4=16$ possible tetrahedra, as shown in Fig.~\ref{fig:diagram}(e): 
six two-in two-out, four three-in one-out, four one-in three-out, one all-in, and one all-out type.
In this complicated structure, the mean-field from the neighboring spins is spatially inhomogeneous; 
namely, the mean-field from NN spins cancels to zero only at the sites belonging to all-in and all-out tetrahedra, while that at other sites has the same nonzero magnitude. 
This point will be discussed later in relation to charge disproportionation and its switching by magnetic field. 

We also calculated the phase diagram at $T=0$ by a variational calculation with limiting the trial states to the four
ordered states appearing in the MC simulation~\cite{note_gsdiagram}. 
The result is shown at the bottom axis of Fig.~\ref{fig:diagram}(a). 
All the four phases appear in the corresponding density
regions where the MC result shows their instabilities.
The phase transitions between different magnetic phases are all first order at
$T=0$, and hence, they are accompanied by a jump of $n_e$, that is
an electronic phase separation (PS)~\cite{note_ps_in_mc}. 

Let us discuss the phase diagram from the viewpoint of the RKKY interactions. 
Figure~\ref{fig:diagram}(b) shows the nearest-, second-, and third-neighbor components of the RKKY interaction calculated by using
the eigenstates for the $J=0$ tight-binding model.
In the lowest density region $n_e \lesssim 0.05$, all three components are positive, namely FM, because of the small Fermi surface.
This is consistent with the ice-ferro order [Fig.~\ref{fig:diagram}(c)]. 
By increasing the electron density above $0.05$, the second- and third-neighbor RKKY interactions change their signs to be AFM,
but the NN interaction remains dominantly FM.
The ice-$(0,0,2\pi)$ order is stabilized as a compromise of these interactions [Fig.~\ref{fig:diagram}(d)].
With further increase of the electron density, the NN RKKY interaction also changes its sign at $n_e \sim 0.20$.
The 32-sublattice order emerges in this region where the NN interaction becomes irrelevant; indeed, it is characterized as AFM
ordering of the second-neighbor spins along the chains, as mentioned above.
In the higher density region $n_e \gtrsim 0.22$, the NN interaction becomes AFM, and the all-in/all-out order is stabilized.

It should be stressed that, although the sequence of magnetic orderings is qualitatively understood,
it is still hard to tell the phase diagram only from the RKKY interactions. 
In particular, the complicated 32-sublattice order is difficult to be predicted by
the simple RKKY analysis~\cite{Ikeda2008}. 
The critical temperature of each phase is also difficult to be predicted;
for instance, although the magnitude of NN ferromagnetic RKKY interaction at
$n_e \sim 0.1$ is comparable to that of the antiferromagnetic one at $n_e \sim 0.3$, 
$T_c$ is much lower in the former region compared to the latter~\cite{Udagawa2012}.
The unbiased MC simulation explicitly taking account of itinerant electron degree of
freedom is crucial to identify the phase diagram and $T_c$.

\begin{figure}
   \begin{center}
   \includegraphics[width=3.20in]{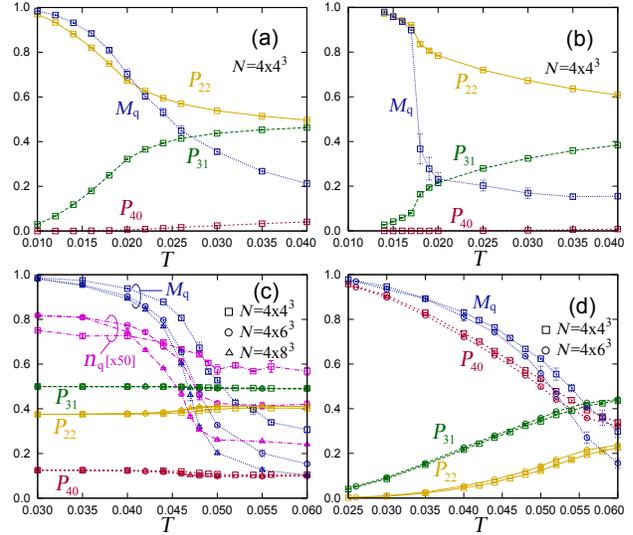}
   \end{center}
   \caption{ (color online).
   $T$ dependences of the order parameter $M_{\bf q}$ and the ratios of tetrahedra with different spin configurations, $P_{22}$, $P_{31}$, and $P_{40}$, at $J=2$.  The data are calculated at a fixed chemical potential $\mu$: (a) $\mu = -5.9$ [corresponding to $n_e = 0.030(2)$], (b) $\mu = -4.8$ [$n_e = 0.099(3)$], (c) $\mu = -3.4$ [$n_e = 0.195(1)$], and (d) $\mu = -2.4$ [$n_e = 0.264(3)$].
   The magnetic ordering wave vectors for each phase are (a),(d) ${\bf q}=0$, (b) $(0,0,2\pi)$, and (c) $(\pi,\pi,\pi)$.
   In (c), the charge disproportionation $n_{\bf q}$ with ${\bf q}=(0,0,2\pi)$ is also plotted.
   }
   \label{fig:structure}
\end{figure}

Next, let us look into $T$ dependences of the order parameters and short-range spin correlations within tetrahedra.
The local spin correlations are measured by calculating the ratio of all-in or all-out ($P_{40}$), three-in one-out or one-in three-out ($P_{31}$), and two-in two-out ($P_{22}$) tetrahedra.
Figure~\ref{fig:structure}(a) shows the results in the ice-ferro region.
The results show an enhancement of the ice-rule local correlation $P_{22}$ prior to phase transition at $T_c \simeq 0.023$, while
 $P_{40}$ and $P_{31}$ are strongly suppressed.
$P_{22}$ becomes dominant also in the ice-$(0,0,2\pi)$ phase region, as shown in Fig.~\ref{fig:structure}(b)~\cite{note_order_of_transition}.
On the other hand, the results for all-in/all-out region show contrastive behavior [Fig.~\ref{fig:structure}(d)], in which the all-in/all-out correlation $P_{40}$ is enhanced prior to the phase transition at $T_c \simeq 0.051$.
In sharp contrast, for the 32-sublattice order, the local correlation parameters show little $T$ dependences, even in the critical region near $T_c \simeq 0.047$.
This implies that the ordering is not driven by NN spin correlations, as discussed above.

Furthermore, the 32-sublattice magnetic order exhibits a concomitant charge disproportionation.
In Fig.~\ref{fig:structure}(c), we plot the charge disproportionation defined by $n_{\bf q} = [N^{(\alpha)}_{{\bf q} \ {\rm max}}/N_{\rm t}]^{1/2}$.
Here, $N^{(\alpha)}_{\bf q} = \sum_{i, j \in \alpha} \langle n_i n_j \rangle \exp\left[ {\rm i} {\bf q}\cdot\left({{\bf r}_i - {\bf r}_j}\right)\right] / N_{\rm t}$ is the charge structure factor and $n_i = \sum_\sigma \langle c_{i\sigma}^\dagger c_{i\sigma} \rangle$ is the local electron density at each site.
The result clearly indicates the emergence of concomitant charge disproportionation below $T_c$
(the temperature of the inflection point of $n_{\bf q}$ agrees with that of $M_{\bf q}$).
The wave number is characterized by ${\bf q}=(0,0,2\pi)$; the local electron density is higher at the sites belonging to the all-in/all-out tetrahedra compared to the other sites.
The pattern appears to be related with the inhomogeneity of the mean-fields from NN spins mentioned above.
We note that all the other phases in Fig.~\ref{fig:diagram} are charge uniform, and the 32-sublattice order is the only phase showing a concomitant charge disproportionation~\cite{note_cdw}.

\begin{figure}
   \begin{center}
   \includegraphics[width=3.20in]{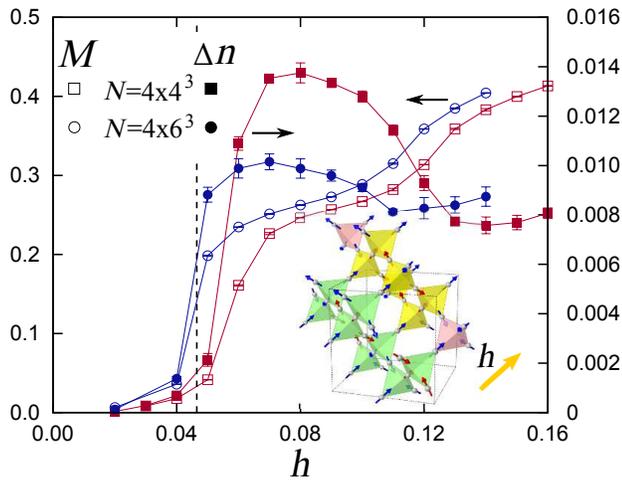}
   \end{center}
   \caption{(color online).
   Magnetization $M$ and charge disproportionation $\Delta n$ under the external magnetic field $h$ along the $[111]$ direction at $\mu=-3.4$ and $T=0.025$.
   The vertical dashed line is an estimate of the critical field at $T=0$, $h_c \simeq 0.0465$, obtained by comparing the ground state energy in a sufficiently large system.
   The inset shows a schematic picture of the magnetic order in the plateau state at $M \simeq 1/4$. 
   }
   \label{fig:magproc}
\end{figure}

The peculiar charge disproportionation can be switched by external magnetic field.
Figure~\ref{fig:magproc} shows the magnetization $M = |\sum_i {\bf S}_i|/N$ and charge disproportionation
under the magnetic field $\bf h$ applied along the $[111]$ direction at $T=0.025$.
Here, the charge disproportionation $\Delta n$ is defined by the difference of the electron densities
between the kagome and triangular planes perpendicular to the $[111]$ direction.
For simplicity, we apply the magnetic field only to the localized Ising spins.
Under weak magnetic field up to $h \sim 0.04$, the 32-sublattice order at $h=0$ remains robust and $M$ stays almost zero.
For higher magnetic field, $M$ increases abruptly, indicating a first order transition to a different phase,
which is characterized by the half magnetization plateau $M \simeq 1/4$.  (The full saturation is $M=1/2$ in the present model.)
This plateau state remains stable up to $h \sim 0.1$. 
The magnetic structure of the plateau phase is obtained by aligning all the spins on the triangular layers to
the field direction (see the inset of Fig.~\ref{fig:magproc}).

Interestingly, $\Delta n$ also changes from almost zero to nonzero abruptly at the critical field, namely,
the charge disproportionation is simultaneously switched to the one along the [111] direction~\cite{note_field}.
The switching may be understood as follows.
As mentioned above, the charge disproportionation with ${\bf q}=(0,0,2\pi)$ in the zero-field state appears to be dominantly driven by
the inhomogeneity of the NN spin correlations. 
In contrast, in the plateau state in applied magnetic field, the NN mean-fields have uniform, nonzero magnitude at all the sites.
Instead, the mean-fields from second-neighbor spins along the chains are different 
between the triangular and kagome sites, which may lead to the [111] charge disproportionation in the plateau state.

To summarize, we have numerically investigated the Kondo lattice model with Ising spins with $\langle 111\rangle$ anisotropy
on a pyrochlore lattice, which potentially describes metallic pyrochlore oxides. 
By the state-of-the-art Monte Carlo simulation with taking account of the itinerant electron degree of freedom,
we found the 32-sublattice ordered phase in the competing region between the two-in two-out and all-in/all-out phases.
This phase exhibits a charge disproportionation concomitant with the magnetic order. 
The spin and charge pattern can be switched by external magnetic field to the
different one which shows a half-magnetization plateau. 
Our result demonstrates that the spin-charge coupling on the frustrated lattice induces
richer behaviors than in the localized spin systems.
It will stimulate further study of not only magnetic but also electronic and transport
properties induced by the spin-charge interplay in frustrated itinerant electron systems.

\begin{acknowledgment}
The authors thank C. D. Batista, H. Kawamura, and T. Misawa for fruitful discussions and helpful comments.
Part of the calculations were performed on the Supercomputer Center, Institute for Solid State Physics, University of Tokyo.
H.I. is supported by Grant-in-Aid for JSPS Fellows.
This research was supported by KAKENHI (No.19052008, 21340090, 21740242, 22540372, 24340076, and 24740221), Global COE Program
``the Physical Sciences Frontier", the Strategic Programs for Innovative Research (SPIRE), MEXT, and
the Computational Materials Science Initiative (CMSI), Japan.
\end{acknowledgment}


\begin{thebibliography}{99}
\bibitem{note_Tc}                  As the current model is a three-dimensional Ising
                                   spin system with effective long-range interactions
				   induced by the kinetic motion of itinerant
				   electrons, it is in general expected to show a
				   long-range order at $T=0$ and a phase transition at
				   finite $T_c$.
\bibitem{note_gray}                The closed symbols at $n_e \sim 0.05$ and $0.15$ indicate phase transitions to other complex magnetic
                                   orderings, whose magnetic unit cells reach the system size; we need larger system sizes to confirm the
				   orderings.
\bibitem{note_wavenumber}          In the fcc primitive unit cell, the ordering vectors are given by $(\pi,\pi,\pi)$, $(\pi,0,0)$,
                                   $(0,\pi,0)$, and $(0,0,\pi)$ for the four sublattices.
\bibitem{note_gsdiagram}           The calculations were done in the grand canonical ensemble by comparing the grand potential
                                   $\Omega = \langle H - \mu \hat{N_c} \rangle$ while varying the chemical potential $\mu$.
				   The evaluation of $\Omega$ was done by numerical diagonalization using $12^3$ superlattices of
				   $4\times 4^3$ sites.
\bibitem{note_ps_in_mc}            The sign of PS is also observed in MC data at low
                                   $T$ as a steep change of $n_e$ with respect to $\mu$,
				   while the PS boundary at finite $T$ is not shown as the
				   precise determination is highly computationally demanding.
\bibitem{note_field}               Although there is a finite size effect, the data for $N=4 \times 6^3$ show the transition to the plateau state
                                   very close to $h_c$ estimated at $T=0$ in the thermodynamic limit. In the high field region above the plateau
				   state, however, we could not obtain converged results due to the poor convergence of PEM.
\end{thebibliography}
\end{document}